\documentclass[]{elsart}
\usepackage{psfig,latexsym}

\newcommand{\lsed}{\mbox{$\ell_{\mbox{\scriptsize SED}}$}}

\begin{document}

\begin{frontmatter}
\title{Evaporation and Step Edge Diffusion in MBE}
\thanks[me]{Corresponding author. FAX:+49-931-888-4604; e-mail:
  schinzer@physik.uni-wuerzburg.de}
\author[TP3]{S. Schinzer\thanksref{me}},
\author[EP2]{M. Sokolowski},
\author[TP3]{M. Biehl},
and \author[TP3]{W. Kinzel}
\address[TP3]{Institut f\"ur Theoretische Physik, TP III }
\address[EP2]{Institut f\"ur Experimentelle Physik, EP II;\\
Universit\"at W\"urzburg, Am Hubland, D-97074 W\"urzburg, Germany}

\begin{abstract}
Using kinetic Monte--Carlo simulations of a Solid--on--Solid model
we investigate the influence of step edge diffusion (SED) and
evaporation on Molecular Beam Epitaxy (MBE). Based on these
investigations we propose two strategies to optimize MBE--growth.  The
strategies are applicable in different growth regimes: during
layer--by--layer growth one can reduce the desorption rate using a
pulsed flux. In three--dimensional (3D) growth the SED can help to
grow large, smooth structures. For this purpose the flux has to be
reduced with time according to a power law.
\end{abstract}

\date{31. August 1998}
\end{frontmatter}

The growth of high quality compound semiconductors is of great
technological importance. In order to optimize growth a detailed
knowledge of microscopic processes is very important. We will follow
this line of thought and will report on two new approaches of growth
which will be presented in more detail in a forthcoming publication.
We will exploit macroscopic effects of two
distinct microscopic mechanisms. The term ``microscopic'' refers to
events at the atomic scale: e.g. a single diffusion step of an adatom or
evaporation of an atom. These are the ingredients of the computer
model. This is contrasted to the term
``macroscopic'' for effects which are typically measurable in
experiments: e.g. the overall mass desorption, form and distribution
of three--dimensional structures, or the growth rate. 

The model used here is the conventional solid--on--solid model on a
simple cubic lattice. All simulations will start on a singular
surface. The system size will be at least $300 \times 300$ lattice
constants with periodic boundary conditions. Even though the model
includes only one species of 
particles it has been successfully applied to reproduce quantitatively
RHEED-oscillations during the growth of GaAs(001) \cite{sv93}. In this
paper we want to study qualitative effects rather than to derive a
model for a specific material. For this purpose we set the vibration
frequency to $\nu_0=10^{-12} s^{-1}$ which is of the order of typical
Debye frequencies. The activation energy for the different microscopic
processes is parameterized in a simple manner. We choose a barrier for
diffusion of $E_B=0.9$ eV, $E_D=1.1$ eV for desorption and at step edges an
additional Ehrlich--Schwoebel barrier $E_S$ of 0.1 eV. Each in--plane
neighbour contributes the binding energy of $E_N=0.25$ eV.
This set of parameters is loosely connected to CdTe(001) and 
reproduces some features of sublimation \cite{sk98,nsk99} and annealing.

\begin{figure}
\psfig{file=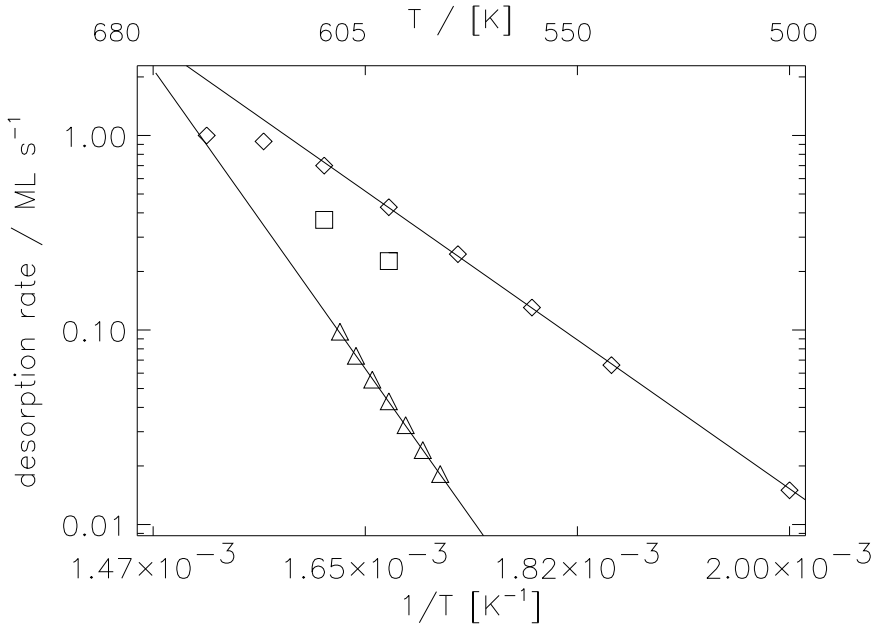,width=12cm}
\caption{The diamonds ($\Diamond$)represent measured desorption rates
  during growth. At high
  temperatures the desorption rate saturates and equals the flux of
  impinging particles (1 ML/s). The desorption rate is considerably
  higher than the sublimation rate of the same model
  ($\bigtriangleup$, sublimation rate is multiplied by ten).  Using
  the proposed ``flush-technique'' we obtain lower desorption rates
  ($\Box$). We used a mean flux of 1 ML/s - a constant flux of 0.77
  ML/s plus additional 0.23 ML in 0.003 s at the beginning of each
  second. 
} \end{figure}

Two of the most important compound semiconductors
decrease their growth rate with increasing temperature in
MBE--growth (CdTe(001) \cite{umm91,ac95,blw95} and GaAs (001)
\cite{tna97}). The desorption rate was found to be activated
with a low activation energy. This should be contrasted to
sublimation with considerably higher activation energies
(c.f. \cite{nst98} for CdTe(001) ). The same qualitative difference in
activation energies is observed in simulations 
\cite{pp98,sk98}. As can be seen in fig. (1) desorption
during growth (flux of arriving particles F = 1 ML/s) is an
activated process with 0.9 eV which is considerably lower than
$E_D$. This is due to a negative contribution of $E_B$ to the
effective energy: high diffusion barriers reduce the diffusion length
more than the typical island separation. Hence, the desorption rate
increases. On the contrary, sublimation (F = 0 ML/s)
shows an activation with 1.7 eV. Here, the
freely diffusing adatoms (which evaporate easily) must be created
through the detachment from steps or the creation of vacancies. Thus
the sublimation energy is much higher than $E_D$.

Our findings imply a simple way to reduce desorption during growth:
short flushes 
of particles at the beginning of each monolayer result in a great
density of islands. Afterwards with a low flux the particles most likely
hit islands to stick to. This reduces the adatom density and hence
yields a low overall desorption
rate. As can be seen in fig. (1) such simulations yield desorption
rates reduced by a factor 2. 
In addition this method leads to prolonged layer--by--layer
growth as will be shown in a forthcoming publication.

Quite generally, layer--by--layer growth is not attainable forever
\cite{kbk97}. 
To optimize MBE--growth it could thus be advantageous to study the
growth of 3D--structures. Using a simplified model \cite{bks98} we
have found a strong influence of SED on the properties of the growing
surface. Clearly, a strong SED leads to structures with smooth step
edges. However, a second consequence of strong SED is a fast
coarsening process. If the structures are becoming too large compared
to the SED--length \lsed\ the coarsening process slows down and the 
step edges become rough. Hence, MBE--growth always drives the surface
in the latter regime with comparably weak SED.
This can be avoided by {\it e.g.} reducing the flux during growth
which in turn increases \lsed. The flux--dependence of the
typical extension of mounds as well as of the SED are power laws in
time. Hence the flux should be reduced according to a power law 
$F(t) \propto 1/t^\omega$. A detailed calculation and determination of
$\omega$ will be published in a forthcoming publication.

\begin{figure}[t]
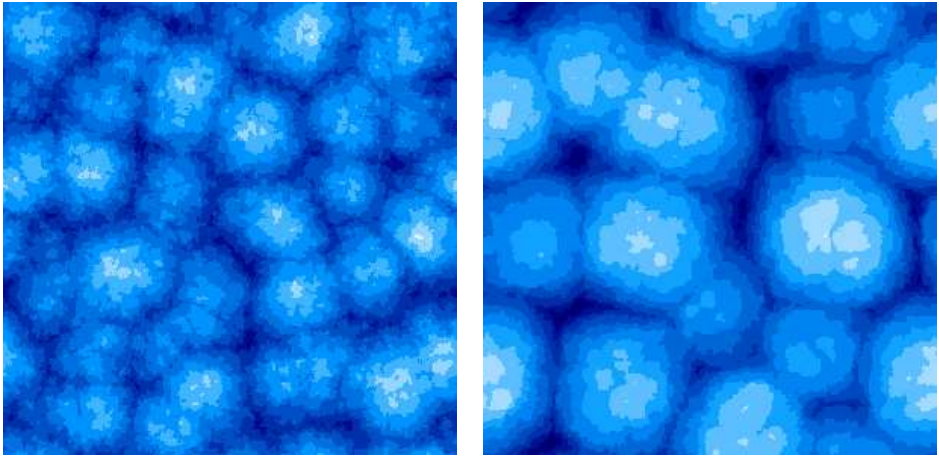

\parbox[b]{3.7cm}{\psfig{file=surf37.ceps,width=6cm}}
\hspace{0.1cm}
\parbox[b]{3.7cm}{\psfig{file=surf90.ceps,width=6cm}}
\caption{                      \label{stat}
Snapshots of surfaces after the deposition of $\langle h
\rangle = 300$ ML. Both simulations started with a flux $F_0 = 1$
ML/s. The right surface corresponds to the simulation
where the flux was reduced during growth according to $F = F_0 / (t/10
s)^{0.1}$ after the deposition of 10 ML's. The system size is $300
\times 300$ in both cases.
}
\end{figure}

We applied this strategy to the growth of the model at $560 K$ which
is a reasonable growth temperature for CdTe(001). For
a first qualitative investigation we choose $\omega = 0.1$. To
prevent the inference of evaporation we inhibited this
process. However, we checked that even with desorption, the strategy
is still applicable and useful. 
After the deposition of 300 ML under constant flux the 
structures are small and of irregular shape. With the adaption of the flux
(in the end $F = 0.11$ ML/s) the
structures are considerably larger (SED assisted coarsening) and the
steps are smoother. 

To summarize, we have shown how the detailed knowledge of underlying
microscopic (atomistic) processes can lead to an optimization of
MBE--growth. If desorption is present a flush--technique can decrease
the desorption rate during layer--by--layer growth.
Besides the improved growth rates such experiments
allow to decide whether physisorption (as speculated in \cite{blw95})
or the above mechanism is dominant.
To obtain larger structures in conventional 3D--growth one just has
to grow for longer times. The step edges will become smooth due to the
equilibration after growth stops. However, during growth a larger
probability for the creation of vacancies or other crystal defects will
be present. After growth stops these faults cannot be
eliminated. This is the point where our second strategy improves the
growth. The step edge diffusion remains always strong
enough to maintain smooth step edges. 

\begin{center}
\bf * * * \end{center}

This work is supported by the Deutsche Forschungsgemeinschaft through
SFB 410.

\bibliography{/users1/schinzer/Arbeit/Literatur,/users1/schinzer/Arbeit/NeueLit,/users1/schinzer/Arbeit/Preprints}
\bibliographystyle{unsrt}

\end{document}